\documentstyle{article}

\begin{document}

\title{Breakdown of the linear approximation in the perturbative
analysis of heat conduction in relativistic systems}

\author{L. Herrera\thanks{On leave from Departamento de F\'\i sica, Facultad
de Ciencias, Universidad 
Central de Venezuela, Caracas, Venezuela and Centro de Astrof\'\i sica
Te\'orica, M\'erida, Venezuela.}, A. Di Prisco$^{*}$\\
\'Area de F\'{\i}sica Te\'orica,Facultad de
Ciencias,\\ Universidad de Salamanca, 37008 Salamanca, Espa\~na.\\
\\ and \and J. Mart\'{\i}nez\\
Grupo de F\'{\i}sica Estad\'{\i}stica, Departamento
de F\'\i sica,\\
Universidad Aut\'onoma de Barcelona,\\
08193 Bellaterra, Barcelona, Espa\~na.}

\maketitle

\begin{abstract}
We analyze the effects of thermal conduction in a 
relativistic fluid just
after its departure from spherical symmetry, on a time scale 
of the order of
relaxation time. Using first order perturbation theory, it 
is shown that, as
in spherical systems, at a critical point the effective inertial mass
density of a fluid element vanishes and becomes negative beyond that
point.  The impact of this effect on the reliability of causality
conditions is discussed.
\end{abstract}

\section{Introduction}

The behaviour of dissipative systems at the very moment when they
depart from hydrostatic equilibrium has been recently studied \cite
{Heetal97,HeMa971,HeMa972}.

As result of these works, it appears that a parameter formed by a
specific combination of relaxation time, temperature, proper energy
density and pressure, may critically affect the evolution of the
object.

Specifically, by means of first order perturbation theory, it was
shown that in the equation of motion of any fluid element, the
inertial mass density term is multiplied by a factor which vanish for
a given value of the parameter $\alpha$ defined below in equation
(\ref{alfa}) (critical point) and changes of sign beyond that value.
Nevertheless, even though self-gravitating systems seem to become
more unstable as the above mentioned parameter grows \cite{HeDiP}, an
exact description of the evolution
\cite{HeMapre} does not indicate any anomalous behaviour of the system at or
beyond the critical point.  Thus, first order perturbation theory
seems to be not reliable close or beyond the critical point.

Although causality conditions prevent, in some cases (pure shear or
bulk viscosity \cite{HeMa972}), reaching the critical point. This may
not be the case in the most general situation (heat conduction plus
viscosity \cite {HeMa972} or even in the non-viscous case
\cite{HeMa971}). It is worth mentioning that causality conditions
were found using a first order perturbation method \cite{HiLi83},
thus, in some cases these conditions must be revised.

However in all these works it has been assumed that spherical
symmetry is preserved. It is therefore pertinent to ask if the
appearance of the critical point is closely related to that kind of
symmetry (spherical) or if it represents a general feature of
dissipative systems.

It is the purpose of this work to provide an answer to the question
above by considering a system which, although initially spherically
symmetric, is submitted to perturbations deviating it from spherical
symmetry.

By examining the equation of motion of an arbitrary fluid element
along the ``meridional'' direction, it is shown that a critical point
also appears in this case, indicating thereby the independence of
this effect from the spherical symmetry.

The paper is organized as follows. In the next section the field
equations and conventions are presented. In section 3 we briefly
present the equation for the heat conduction and write down
explicitly the $\theta $-component of this equation. In section 4 the
full system of equations is evaluated at the time when the object
starts to depart from spherical symmetry. Finally, a discussion of
results is given in the last section.

In the whole text, a bar over a quantity denote that this one has
been measured by a Minkowskian observer.  If the symbol over a given
quantity is a tilde, then this one is evaluated after the
perturbation.

\section{The field equations and conventions}

We shall consider axially and reflection symmetric distributions of
fluid dissipating energy through a heat flux vector. In null
coordinates the metric is given by the standard Bondi expression
\cite{Bo62}

\begin{equation}
\begin{array}{lll}
ds^{2} & = & \left(\frac{V}{r}e^{2\beta}-U^{2}r^{2}e^{2\gamma}
\right)du^{2}+2e^{2 \beta}du\,dr \\
&  &
+2Ur^{2}e^{2\gamma}du\,d\theta-r^{2}\left(e^{2\gamma}d\theta^{2}+
e^{-2\gamma}\sin^{2}\theta\: d\phi^{2}\right),
\end{array}
\label{bo}
\end{equation}

where $V$, $\beta$, $U$ and $\gamma$ are functions of $u$, $r$ and
$\theta$.

We number the coordinates
$x^{0,1,2,3}=u,r,\theta,\phi,\;$respectively. $\;u$ is a time-like
coordinate such that $u=constant$ defines a null surface. In flat
space-time this surface coincides with the null light cone open to
the future. $r$ is a null coordinate ($g_{rr}=0$) and $\theta$ and
$\phi$ are two angle coordinates (see \cite{Bo62} for details).

In order to guarantee the regularity of the metric within the fluid
distribuition at $r=0$, we impose the conditions\cite{IsWeWi83}

\begin{equation}
\begin{array}{cc}
V=r+O(r^{3}); & \beta=O(r^{2}); \\ U=O(r); & \gamma=O(r^{2}).
\end{array}
\label{cois}
\end{equation}

In this case, the energy-momentum tensor takes the usual form

\begin{equation}
T_{\mu\nu}=(\rho+p)v_{\mu}v_{\nu}-pg_{\mu\nu} + q_\mu v_\nu + q_\nu
v_\mu
\label{T}
\end{equation}

where $\rho$, $p$ and $q_\mu$ denote proper energy density, pressure
and the heat flow vector, respectively, and the four-velocity
$v^{\mu}$ satifies the conditions

\begin{equation}
v^{\mu}v_{\mu}=1 \qquad \,;\, \qquad q^\mu v_\mu=0  \label{cov}
\end{equation}

Since reflection symmetry is preserved, we have

\begin{equation}
v^{3}=v_{3}=0 \qquad \,;\, \qquad q^{3}=q_{3}=0  \label{sire}
\end{equation}

We shall now write the field equations according to the scheme
presented by Isaacson {\it et al.} \cite{IsWeWi83} (see also
\cite{ChWi73}). Thus, there are three main equations which only
contain derivatives with respect to $r$ and $\theta$. In our case the
corresponding expressions are given by

\begin{equation}
2\pi r\left[ \left( \rho +p\right) v_1+2q_1\right] v_1=\beta
_{,1}-\frac 12r\left( \gamma _{,1}\right) ^2  \label{meq1}
\end{equation}

\begin{eqnarray}
16\pi r^2\left[ \left( \rho +p\right) v_1v_2+q_1v_2+q_2v_1\right]
=\left[ r^4e^{2(\gamma -\beta )}U_{,1}\right] _{,1}-  \nonumber \\
2r^2\left[ r^2\left( r^{-2}\beta \right) _{,12}-\sin ^{-2}\theta
\left(
\gamma \sin ^2\theta \right) _{,12} + 2\gamma _{,1}\gamma _{,2}\right]
\label{meq2}
\end{eqnarray}

\begin{eqnarray}
- 8\pi r^2 e^{2\beta } \left[ \rho -p+r^{-2}e^{-2\gamma }\left(
\left[ \rho +p\right] (v_2)^2+2q_2v_2\right) \right] =  \nonumber \\
2V_{,1}+\frac{r^4}4e^{2(\gamma -\beta )}(U_{,1})^2-\left( r^2\sin
\theta
\right) ^{-1}(r^4U\sin \theta )_{,12}+ 2e^{2(\beta -\gamma )}\times
\nonumber \\ 
\left[ -1+\left( \sin \theta \right) ^{-1}(\beta
_{,2}\sin \theta )_{,2}-\gamma _{,22}-3\gamma _{,2}\cot \theta
+(\beta _{,2})^2+2\gamma _{,2}(\gamma _{,2}-\beta _{,2})\right]
\label{meq3}
\end{eqnarray}

where a comma denotes partial derivative. Following the
Bondi-Isaacson scheme we shall assume that variables $\rho$, $p$,
$v_{1}$, $v_{2}$, $q_{1}$ , $q_{2}$ and $\gamma$ are known on a
given initial null hypersurface ($ u=u_{0}=constant$), which we
shall call the ``initial cone''.

At $u=u_{0}$ perturbations are introduced into the system, forcing it
to deviate from spherical symmetry. We shall evaluate the system
immediately after perturbing it, on a null surface ($u=\tilde{u}
=constant$) such that $
\tilde{u}-u_{0}$ is much smaller than the typical time required for $v_{1}$,
$v_{2}$, $q_{1}$, $q_{2}$ and $\gamma$ to change significantly (where
``significantly'' refers to terms linear in the perturbation, or
larger). In other words, on the surface $u=\tilde{u}$, which we shall
call the ``evaluation cone'', the components $v_{1}$ and $v_{2}$ of
the four-velocity, $q_{1}$ and $q_{2}$ of the heat flow vector and
the metric function $\gamma$ have the same magnitude as on the
initial cone (up to cuadratic and higher terms in the perturbation
parameter).

The ``standard equation'' which provides an expression for
$\gamma_{,0}$ reads

\begin{eqnarray}
4r\left( r\gamma \right) _{,01}&=&\left[ 2r\gamma _{,1}V-r^2\left(
2\gamma _{,2}U+U_{,2}-U\cot \theta \right) \right] _{,1}-2r^2\sin
^{-1}\theta \left(
\gamma _{,1}U\sin \theta \right) _{,2}  \nonumber \\
 & &+\frac{r^4}2e^{2(\gamma -\beta )}(U_{,1})^2+2e^{2(\beta
-\gamma)}\times
\nonumber \\
 & &\left[\left( \beta _{,2}\right) ^2+\beta _{,22}-\beta _{,2}\cot
\theta +4\pi
\left( \left[ \rho +p\right] (v_2)^2+2q_2v_2\right) \right]
\label{seq}
\end{eqnarray}

Next, in order to give physical meaning to quantities appearing in
(\ref{T}), taking into account the axial symmetry of the case
considered, we shall develop a procedure similar to that used by
Bondi \cite{Bo64} in his study of non-static spherically symmetric
sources. Thus, let us introduce purelly local Minkowski coordinates
($\overline{x}^{0}, \overline{x}^{1}, \overline{x}^{2},
\overline{x}^{3}$), defined by \cite{HeVa97}

\begin{equation}
d\overline{x}^{0}=Adu+Bdr+Cd\theta  \label{x0}
\end{equation}

\begin{equation}
d\overline{x}^{1}=Bdr+Cd\theta  \label{x1}
\end{equation}

\begin{equation}
d\overline{x}^2=Fd\theta  \label{x2}
\end{equation}

\begin{equation}
d\overline{x}^{3}=Gd\phi  \label{x3}
\end{equation}

 where cofficients $A$, $B$, $C$, $F$ and $G$ are given by

\begin{equation}
A=\left(\frac{V}{r}e^{2\beta}-U^{2}r^{2}e^{2\gamma}\right)^{\frac{1}{2}}
\label{A}
\end{equation}

\begin{equation}
B=e^{2\beta}\left(\frac{V}{r}e^{2\beta}-U^{2}r^{2}e^{2\gamma}\right)^{-\frac
{1}{2}}  \label{B}
\end{equation}

\begin{equation}
C=Ur^{2}e^{2\gamma}\left(\frac{V}{r}e^{2\beta}-U^{2}r^{2}e^{2\gamma}
\right)^{ -\frac{1}{2}}  \label{C}
\end{equation}

\begin{equation}
F=re^{\gamma}  \label{F}
\end{equation}

\begin{equation}
G=re^{-\gamma}\sin\theta  \label{G}
\end{equation}

so that,

\begin{equation}
ds^{2}=(d\overline{x}^{0})^{2}-(d\overline{x}^{1})^{2}-(d\overline{x}
^{2})^{2} -(d\overline{x}^{3})^{2}
\end{equation}

Now, for the Minkowski observer defined above the fluid moves in such
a way that the four-velocity vector of the fluid is given by

\begin{equation}
\overline{v}^{\mu}=\lambda\left(1,\omega_{1},\omega_{2},0\right),
\label{vbar}
\end{equation}

where $\omega_{1}$ and $\omega_{2}$ are the components of the
velocity along the $\overline{x}^{1}$ and $\overline{x}^{2}$ axes,
respectively (as measured by our locally Minkowski observer), and

\begin{equation}
\lambda=\frac{1}{\sqrt{1-\omega_{1}^{2}-\omega_{2}^{2}}}  \label{lam}
\end{equation}

Thus, a locally Minkowski observer comoving with the fluid is one for
which $
\omega_{1}=\omega_{2}=0$.

Now, from (\ref{vbar}) and (\ref{x0})--(\ref{x3}) we can obtain
expressions for the four-velocity components in the original Bondi
coordinates. In terms of $\omega_{1}$ and $\omega_{2}$, they are

\begin{equation}
v^{\mu}= \lambda
\left[\frac{1-\omega_{1}}{A},\frac{1}{B}\left(\omega_{1}-
\frac{C\omega_{2}}{F} \right),\frac{\omega_{2}}{F},0\right]  \label{vlam}
\end{equation}

or, for the covariant components,

\begin{equation}
v_{0}=\lambda\left(\frac{V}{r}e^{2\beta}-U^{2}r^{2}e^{2\gamma}\right)^
{\frac{1}{2}}  \label{v0}
\end{equation}

\begin{equation}
v_{1}=\frac{\lambda e^{2\beta}
\left(1-\omega_{1}\right)}{\left(\frac{V}{r} e^{2\beta}
-U^{2}r^{2}e^{2\gamma}\right)^ {\frac{1}{2}}}  \label{v1}
\end{equation}

\begin{equation}
v_{2}=\lambda\left[\frac{Ur^{2}e^{2\gamma}\left(1-\omega_{1}\right)}
{\left(
\frac{V}{r}e^{2\beta}-U^{2}r^{2}e^{2\gamma}\right)^{1/2}}-re^{\gamma}
\omega_{2}\right]  \label{v2}
\end{equation}

\begin{equation}
v_{3}=0  \label{v3}
\end{equation}

It is worth noticing that the Minkowski observer will say that the
fluid is at rest when $\omega_1 = \omega_2 = 0$. Also, observe that
$\omega_2$, as it follows from (\ref{x2}), measures the velocity of a
fluid element along the $\theta$ (meridional) coordinate line.

Finally, defining $q$ as

\begin{equation}
q = \sqrt{- q^\mu q_\mu}  \label{q}
\end{equation}

we obtain

\begin{equation}
q_0 = q \lambda \omega_1 \sqrt{\frac{V}{r} e^{2 \beta} - U^2 r^2 e^{2
\gamma} }  \label{q0}
\end{equation}

\begin{equation}
q_1 = - q e^{2 \beta} \left(\frac{V}{r} e^{2 \beta} - U^2 r^2 e^{2
\gamma}\right)^{-1/2} \left(\frac{\lambda \omega_1^2 + \omega_2^2}{
\omega_1^2 + \omega_2^2} - \lambda \omega_1\right)  \label{q1}
\end{equation}

\begin{eqnarray}
q_2 &=& - q r e^\gamma \times  \label{q2} \\ & &\left[U r e^\gamma
\left(\frac{V}{r} e^{2 \beta} - U^2 r^2 e^{2
\gamma}\right)^{-1/2} \left(\frac{\lambda \omega_1^2 + \omega_2^2}{
\omega_1^2 + \omega_2^2} - \lambda \omega_1\right) + \frac{\omega_1 \omega_2
}{\omega_1^2 + \omega_2^2} \left(\lambda - 1\right) \right] \nonumber
\end{eqnarray}

\begin{equation}
q_3 = 0  \label{q3}
\end{equation}

\section{Heat Conduction Equation.}

In the study of star interiors it is usually assumed that the energy
flux of radiation (and thermal conduction) is proportional to the
gradient of temperature (Maxwell-Fourier law or Eckart-Landau in
general relativity).

However, it is well known that the Maxwell-Fourier law for the
radiation flux leads to a parabolic equation (diffusion equation)
which predicts propagation of perturbation with infinite speed (see
\cite{JoPr89}--\cite {Maa} and references therein). This simple fact
is at the origin of the pathologies \cite{HiLi83} found in the
approaches of Eckart \cite{Ec40} and Landau \cite{LaLi} for
relativistic dissipative processes.

To overcome such difficulties, different relativistic theories with
non-vanishing relaxation times have been proposed in the past
\cite{Is}--
\cite{Ca76}. The important point is that all these theories provide a heat
transport equation which is not of Maxwell-Fourier type but of
Cattaneo type
\cite{Cat}, leading thereby to a hyperbolic equation for the propagation of
thermal perturbation.

Accordingly we shall describe the heat transport by means of the
relativistic Israel-Stewart equation \cite{Maa} , which reads

\begin{equation}
\tau \frac{Dq^\alpha}{Ds} + q^\alpha = \kappa P^{\alpha \beta}
\left(T_{,\beta} - T a_\beta\right) - \tau v^\alpha q_\beta a^\beta- \frac{1
}{2} \kappa T^2 \left(\frac{\tau}{\kappa T^2} v^\beta\right)_{;\beta}
q^\alpha  \label{Catrel}
\end{equation}

where $\kappa$, $\tau$, $T$ and $a^\beta$ denote thermal
conductivity, thermal relaxation time, temperature and the components
of the four-acceleration, respectively. Also, $P^{\alpha \beta}$ is
the projector onto the hypersurface orthogonal to the four velocity
$v^\alpha$ and
\[
\frac{Dq^\alpha}{Ds} \equiv v^\beta q^\alpha_{;\beta} \equiv \dot q_\alpha
\]

For the purpose of this work we only need the components of
(\ref{Catrel}) containing $u$-derivatives of $q_2$ and $v_2$. Such
terms only appear in the $\theta$-component of (\ref{Catrel}). Thus
we have

\begin{equation}
\tau \stackrel{.}{q}_\nu P^{\nu 2}+q^2=\kappa P^{\alpha 2}\left( T_\alpha
-Ta_\alpha \right) -\frac 12\kappa T^2\left( \frac \tau {\kappa
T^2}v^\beta
\right) _{;\beta }q^2  \label{teco}
\end{equation}

or simbolically

\begin{equation}
\tau g^{01} v_1 q_{2,0} P^{22} + \kappa T P^{22} g^{01} v_1 v_{2,0} = 
{\cal H}  \label{sim}
\end{equation}

where ${\cal H}$ is a combination of terms not containing $q_{2,0}$
or $ v_{2,0}$.

Alternatively, we may write from (\ref{sim})

\begin{equation}
q_{2,0}=-\frac{\kappa T}\tau v_{2,0}+{\cal I},  \label{q20}
\end{equation}

where ${\cal I}$, as ${\cal H}$, represents a combination of terms
not containing $q_{2,0}$ or $v_{2,0}$.

We shall now get into the central problem of this work.

\section{Thermal conduction and departure from hydrostatic equilibrium}

Let us now consider a fluid distribution which on the initial cone
($u=u_0$) is spherically symmetric. We shall not impose any further
restrictions on it, so that in principle the system may not be in
hydrostatic equilibrium along the radial direction at $u=u_0$.

Next, let us assume that the system is perturbed, as a result of
which it departs from spherical symmetry.

Since the system is initially spherically symmetric it is clear that
on the initial cone the fluid is in equilibrium along the
$\theta$-line of coordinates (i.e. at $u=u_0, \omega_2 = q_2 = 0$ and
$\omega_{2,0} = q_{2,0} = 0$).

For the purpose of this paper (i.e. in order to put in evidence the
existence of a critial point along a non-radial direction) it will
suffice to consider those equations containing $q_{2,0}$ and
$\omega_{2,0}$.

Let us now evaluate our system on the evaluation cone ($u=\widetilde
u$).

As mentioned in the precedent section, the only component of
(\ref{Catrel}) containing terms with $u$-derivatives of $q_2$ or
$\omega_2$ (the $\theta$ -component) leads to eq.(\ref{q20}).

This last equation after perturbation keeps the same form as
(\ref{q20}), where all the quantities are evaluated on the
$u=\widetilde u$ cone, i.e.

\begin{equation}
\widetilde{q}_{2,0}=-\frac{\kappa \widetilde{T}}\tau \widetilde{v}_{2,0}+
\widetilde{{\cal{I}}}.  \label{tq20}
\end{equation}

Next, instead of dealing with the field equations
(\ref{meq1})--(\ref{seq}), it will be more useful to consider the
``conservation'' equations

\begin{equation}
T^\mu_{\nu;\mu} = 0  \label{cons}
\end{equation}

Thus, the $u$-component of (\ref{cons}) gives an expression of the
form

\begin{equation}
\left[\left(\rho + p\right) v_0 + q_0\right] v_{1,0} + \left[\left(\rho +
p\right) v_1 + q_1\right] v_{0,0} + q_{1,0} v_0 + q_{0,0} v_1 = {\cal
J}
\label{ucom}
\end{equation}

where ${\cal J}$ denotes a combination of terms without
$u$-derivatives of matter variables. Observe that (\ref{ucom}) does
not contain terms with $u$ -derivatives of $q_2$ or $\omega_2$. Also
note that all terms containing $
\omega_{2,0}$, appearing from the $u$-derivatives of $\lambda$ in $v_{1,0}$,
$v_{0,0}$, $q_{1,0}$ and $q_{0,0}$ are multiplied by $\omega_2$ and
therefore vanish on the evaluation cone (see
eqs.(\ref{vlam})--(\ref{q3})).

Then, (\ref{ucom}) may be written alternatively as

\begin{equation}
f\left( \omega _{1,0};q_{,0}\right) ={\cal K},  \label{k}
\end{equation}

where ${\cal K}$, as ${\cal J}$, is a combination of terms without
$u$ -derivatives of matter variables.

The $r$-component of (\ref{cons}) yields an expression of the form

\begin{equation}
\left[ \left( \rho +p\right) v_1+q_1\right] v_{1,0}+q_{1,0}v_1={\cal L},
\label{rcom}
\end{equation}

where ${\cal L}$ is again a combination of terms without
$u$-derivatives of matter variables.

As in the precedent case, when evaluating (\ref{rcom}) at $u =
\widetilde u$ , all terms with $\omega_{2,0}$ will vanish since they
appear multiplied by $
\omega_2$, and therefore this later equation may be written as

\begin{equation}
g\left(\omega_{1,0} ; q_{,0}\right) = {\cal M}  \label{M}
\end{equation}

${\cal M}$ denoting another combination of terms without $u$-derivatives of
matter variables.

Finally, let us consider the $\theta$-component of (\ref{cons}), it leads to
an expression of the form

\begin{equation}
\left[\left(\rho + p\right) v_2 + q_2\right] v_{1,0} + \left[\left(\rho +
p\right) v_1 + q_1\right] v_{2,0} + q_{1,0} v_2 + q_{2,0} v_1 = {\cal N}
\label{tcom}
\end{equation}

or, using (\ref{k}) and (\ref{M})

\begin{equation}
\left[\left(\rho + p\right) v_1 + q_1\right] v_{2,0} + q_{2,0} v_1 = {\cal N}
+ h\left({\cal K}, {\cal M}\right)  \label{N+h}
\end{equation}

Since ${\cal N}$ does not contain terms with $u$-derivatives of matter
variables, these later terms are absent on the right hand side of (\ref{N+h}
).

Before perturbation (on the initial cone) (\ref{N+h}) is just an identity 
$0=0$, since the system is spherically symmetric. After perturbation (\ref
{N+h}) keeps the same form, with all quantities evaluated at $u = \widetilde
u$, i.e.

\begin{equation}
\left[\left(\widetilde \rho + \widetilde p\right) \widetilde v_1 +
\widetilde q_1\right] \widetilde v_{2,0} + \widetilde q_{2,0} \widetilde v_1
= \widetilde {{\cal N}} + \widetilde h\left({\cal K}, {\cal M}\right)
\label{N+hti}
\end{equation}

or replacing $\widetilde q_{2,0}$ by (\ref{tq20})

\begin{equation}
\left[ \left( \widetilde{\rho }+\widetilde{p}-\frac{\kappa \widetilde{T}}
\tau \right) \widetilde{v}_1+\widetilde{q}_1\right] \widetilde{v}_{2,0}+
\widetilde{{\cal I}}\widetilde{v}_1=\widetilde{{\cal N}}+\widetilde{h}\left(
{\cal K},{\cal M}\right)   \label{req}
\end{equation}

or, after re-arranging terms

\begin{equation}
\left(\widetilde \rho + \widetilde p - \frac{\kappa \widetilde T}{\tau} +
\frac{\widetilde q_1}{\widetilde v_1}\right) \widetilde v_{2,0} = \frac{
\widetilde {{\cal N}} + \widetilde h\left({\cal K}, {\cal M}\right)}{
\widetilde v_1} - \widetilde {{\cal I}}  \label{arr}
\end{equation}

Next, taking into account that

\begin{equation}
\frac{\widetilde q_1}{\widetilde v_1} = - \widetilde q  \label{-q}
\end{equation}

as it follows from (\ref{v1}) and (\ref{q1}) imposing $\omega _2=0$, and
keeping terms linear in the perturbation, we may write (\ref{arr}) as

\begin{equation}
\left( \rho +p-\frac{\kappa T}\tau -q\right) \widetilde{v}_{2,0}=\frac{
\widetilde{{\cal N}}+\widetilde{h}\left( {\cal K},{\cal M}\right) }{
\widetilde{v}_1}-\widetilde{{\cal I}},  \label{sus}
\end{equation}

It remains to expand $\widetilde{v}_{2,0}$ by using (\ref{v2}). One obtains
terms with $\omega _{2,0}$, terms with $\omega _{1,0}$ and terms not
containing $u$-derivatives of matter variables. The latter may be included
in the right hand of (\ref{sus}). Terms with $\omega _{1,0}$ may also be
transformed into terms not containig $u$-derivatives of matter variables by
virtue of (\ref{k}) and (\ref{M}). Thus, we obtain finally

\begin{equation}
R_{\theta} = - \omega_{2,0} \times \left[1 - \frac{\kappa T}{\tau \left(\rho
+ p - q\right)}\right] \left(\rho + p - q\right)  \label{R}
\end{equation}

The physical meaning of $R_{\theta}$ is quite simple. If the fluid is in
equilibrium along any meridional line then $R_{\theta}=0$, thus $R_{\theta}$
represents the total force on any fluid element in the $\theta$-direction
(observe that $p_{,2}$ enters in $R_{\theta}$).

Therefore (\ref{R}) may be regarded as an equation of the ``Newtonian'' form
\[
Force = mass \times acceleration
\]

where by ``$acceleration$'' we mean the time-like derivative of $\omega_2$.

Thus, if

\begin{equation}
\alpha \equiv \frac{\kappa T}{\tau \left(\rho + p - q\right)} = 1
\label{alfa}
\end{equation}

the effective inertial mass vanishes and any fluid element is out of
hydrostatic equilibrium ($\omega_{2,0} \not= 0$) even though $R_\theta $ is
zero. Note that beyond the critical point ($\alpha >1$), the system exhibits
an anomalous behaviour: If the total force on any fluid element is acting in
the positive $\theta $-direction, then the system tends to accelerate in the
negative $\theta $-direction.

This is the same critical point already found in the spherically symmetric
case \cite{Heetal97,HeMa971,HeMa972}. The only difference with the later
case being the appearance of $q$ in $\alpha$. This is due to the fact that
in the spherically symmetric case the system is initially static ($q=0$) or
slowly evolving ($q$ is small). However in our case, we have considered the
more general situation when the system initially is spherically symmetric
but, although $q_2$ vanishes on the initial cone, $q_1$, and therefore $q$,
is not necessarily zero or a small quantity.

\section{Conclusions}

We have described the departure of a fluid distribution from hydrostatic
equilibrium along the meridional direction, which in its turn implies
departure from spherical symmetry. It has been shown that a critical point,
for which the effective inertial mass density of a fluid element vanishes,
exists. It has been therefore established, that the occurence of such
critical point is independent of spherical symmetry.

As in spherically symmetric systems \cite{Heetal97,HeMa971,HeMa972}, condition
$\alpha =1$ establishes the
upper limit for which first order perturbation theory can be applied. Thus,
close to
(and beyond) this point, causality conditions obtained from the linear
approximation
should be taken with caution.

\section*{Acknowledgment}

One of us (J.M.) would like to express his thanks to the Theoretical Physics 
Group for hospitality at the Physics Department of Universidad de Salamanca. 
This work was partially supported by the Spanish Ministry of Education under
Grant No. PB94-0718

\end{document}